\documentclass[letterpaper]{sig-alternate-10pt}
\usepackage{graphicx}
\usepackage[center]{caption2}

\emergencystretch=\maxdimen
\usepackage[top=0.75in, bottom=1in, left=0.75in, right=0.75in]{geometry}
\usepackage{flushend}
\flushend

\begin{document}
%

\title{ART-GAS: An Adaptive and Real-Time GTS Allocation Scheme for IEEE 802.15.4}

\author{Feng Xia, Ruonan Hao, Yang Cao, Lei Xue\\
School of Software, Dalian University of Technology\\Dalian 116620, China\\
f.xia@acm.org; \{ruonan.h; cao.liz.yang; carmark.dlut\}@gmail.com}



\maketitle
\begin{abstract}
 IEEE 802.15.4 supports a Guaranteed Time Slot (GTS) allocation mechanism for time-critical and delay-sensitive data transmissions in Wireless Personal Area Networks (WPANs). However, the inflexible first-come-first-served GTS allocation policy and the passive deallocation mechanism significantly reduce network efficiency. In this paper, we propose an Adaptive and Real-Time GTS Allocation Scheme (ART-GAS) to provide differentiated services for devices with different priorities, which guarantees data transmissions for time-sensitive and high-traffic devices. The bandwidth utilization in IEEE 802.15.4-based PAN is improved. Simulation results show that our ART-GAS algorithm significantly outperforms the existing GTS mechanism specified in IEEE 802.15.4.
\end{abstract}




\section{Introduction}
Wireless Sensor Networks (WSNs) are attracting growing attention from both academia and industry. The interest is mainly driven by the large amount of WSN applications, including environmental monitoring, industrial sensing and diagnostics, health care, etc. Most of them are developed by using low-rate, short-distance and low-cost wireless technologies. Among the well-known specifications, IEEE 802.15.4 \cite{1}, which was originally designed for Low-Rate Wireless Personal Area Networks (LR-WPANs), has become one of the promising candidates for the fourth-generation wireless network technology \cite{2}.

The standard provides specifications for the Physical Layer (PHY) and the Medium Access Control (MAC) sublayer. Specifically, the MAC design of IEEE 802.15.4 follows the modified Carrier Sense Multiple Access with Collision Avoidance (CSMA/CA) mechanism and the Guaranteed Time Slot (GTS) mechanism. Tremendous works have already analyzed the performance of the IEEE 802.15.4 MAC [3,4]. One of the main design goals of the standard has been energy efficient operation, whereas adaptive and real-time aspects were not a primary concern [5]. Admittedly, The GTS mechanism is designed to support time-critical data transfers generated by repetitive low-latency applications. This mechanism serves to allocate a specific duration within a superframe for data transmissions and guarantee the reliability and performance of data deliveries. Nevertheless, the abuse of dedicated resources might result in the exclusion of other transmissions. This issue is further complicated by the First-Come-First-Served (FCFS) GTS allocation policy [1] and the fixed timer maintained in IEEE 802.15.4 for GTS deallocation.

The performance of the IEEE 802.15.4 protocol has been subject of many research studies recently. These studies include performance analysis of the CSMA/CA protocol [6,7] in Contention Access Period (CAP), and the GTS mechanism [8,9] operating in the Contention Free Period (CFP) is also concerned. Specifically, some interesting algorithms are proposed to improve the performance of GTS allocation mechanism. To support a limited number of GTSs for time-critical and delay-sensitive data transmissions, an optimization-based GTS allocation scheme is proposed in [10]. In [11], Koubaa et al. propose an implicit GTS allocation mechanism (i-GAME) to improve the GTS utilization efficiency. Huang et al. propose an adaptive GTS allocation scheme by considering the low delay and fairness in [12]. In order to maximize the bandwidth utilization, the smaller slot size and offline message scheduling algorithm are proposed in [13] and [14], respectively.

In this paper, we mainly focus on the GTS allocation mechanism in IEEE 802.15.4 and propose an Adaptive and Real-Time GTS allocation Scheme (ART-GAS) for time-critical and delay-sensitive applications. The proposed ART-GAS is developed based on the standard of the IEEE 802.15.4 MAC protocol and completely follows the specification defined in [1] without introducing any extra protocol overhead. Our method is compared with the FCFS strategy adopted by the standard, showing that our solution provides a significant improvement in terms of average delay and fairness performance.

The remainder of this paper is organized as follows: In Section 2, we describe the IEEE 802.15.4 MAC protocol and the standard GTS allocation mechanism. Section 3 defines the problem under investigation and proposes our ART-GAS algorithm to provide adaptive and real-time guarantees for the IEEE 802.15.4. In Section 4, we give simulation setup of ART-GAS and the performance evaluation of our ART-GAS through a myriad of experiments. Finally, Section 5 concludes the paper.

\section{OVERVIEW OF IEEE 802.15.4 MEDIUM ACCESS CONTROL}
IEEE 802.15.4 is a standard for low-rate, low-power and low-cost Personal Area Networks (PANs) [1]. It defines two different channel access methods: beacon-enabled mode and non beacon-enabled mode.

In beacon-enabled mode, beacon frames are periodically sent by the PAN coordinator to identify its PAN and synchronize nodes that are associated with it. The PAN coordinator defines a superframe structure characterized by a Beacon Interval ($BI$) and a Superframe Duration ($SD$). They are determined by Beacon Order ($BO$) and Superframe Order ($SO$) respectively, which are broadcasted by the coordinator via a beacon to all nodes. ~$BI$~ specifies the time between two consecutive beacons, and includes an active period and, optionally an inactive period. The active period, also called superframe, is corresponding to ~$SD$~ and can be divided into 16 equally-sized time slots, during which frame transmissions are allowed. During the inactive period (if it exists), all nodes may enter into a low-power state to save energy. The superframe structure of beacon-enabled mode is depicted in Fig. 1.

\begin{figure}[!t]
\setlength{\abovecaptionskip}{5pt}
\setlength{\belowcaptionskip}{5pt}
\renewcommand{\captionfont}{\bfseries}
\centering
\includegraphics[width=3.2in]{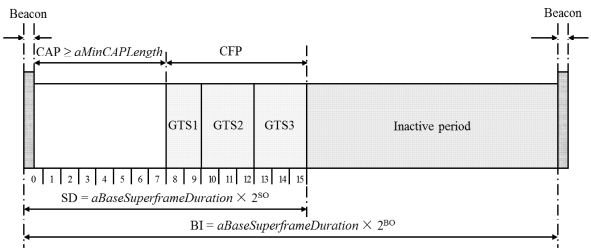}
\centering
\caption{IEEE 802.15.4 superframe structure}
\label{fig_sim}
\end{figure}

~$SD$~ can be further divided into CAP and CFP. The beacon is transmitted by the coordinator at the start of slot 0, and the CAP follows immediately after the beacon. During the CAP, a slotted CSMA/CA algorithm is used for channel access. A node computes its backoff delay based on a random number of backoff periods, and performs two Clear Channel Assessments before accessing the medium.

In the CFP, which is for the use of devices requiring dedicated bandwidth, communication occurs in a TDMA (Time Division Multiple Access) style by using a number of GTSs, pre-assigned to the individual sensor nodes. Whenever a device requires a certain guaranteed bandwidth for transmission, the device sends GTS request command using CSMA/CA during CAP. Upon receiving the request, the coordinator first checks the availability of GTS slots in the current superframe, based on the remaining length of the CAP and the desired length of the requested GTS. The superframe shall have available capacity if the maximum number of GTSs has not been reached and allocating a GTS of the desired length would not reduce the length of the CAP to less than ~$aMinCAPLength$~. Provided there is sufficient capacity in the current superframe, the coordinator determines, based on a FCFS fashion, a device list for GTS allocation in the next superframe, and informs the device about the allocation of slot in the GTS descriptor in the following beacon frame.

GTS deallocation can be performed by the coordinator or by the device itself. For device initialized deallocation, it sends GTS request with characteristic type subfield set to zero using CSMA/CA during CAP. From this point onward, the GTS to be deallocated shall not be used by the device, and its stored characteristics shall be reset. In this way, devices can return the GTS resources by explicitly requesting that the PAN coordinator provide deallocation. However, in most cases, the PAN coordinator has to detect the activities of the devices occupying GTSs and determine when the devices stop using their GTSs. If the coordinator does not receive data from the device in the GTS for at least 2*n super frames, the coordinator will deallocate the GTS with starting slot subfield set to zero in the GTS descriptor field of the beacon frame for that device, where ~$n = 2^{8-BO} for 0 \le BO \le 8$~, and ~$n = 1 for 9 \le BO \le 14$~.

\section{ART-GAS: Adaptive and Real-Time GTS Allocation}
The objective of this section is to propose the ART-GAS for the IEEE 802.15.4 standard using a star topology. Consider, for example, the PAN coordinator collects data from different sensors deployed in the body. Data from these sensors are collected periodically. However, emergency data may be generated randomly and need to be transmitted immediately. Furthermore, the GTS resources should be carefully allocated to needy devices with higher frequencies of sending data and the previously allocated but unused GTSs should be reclaimed in time to achieve network efficiency.

To solve these problems, our ART-GAS adopts a GTS scheduling approach, which is based on the service differentiation mechanism and the GTS allocation mechanism. In the former mechanism, devices are assigned two different kinds of priorities in a dynamic fashion: data-based priority and rate-based priority. Devices which are sending data of greater importance or with real-time requirements are given higher data-based priorities, while devices with higher frequencies of sending data are given higher rate-based priorities. In the latter mechanism, a comprehensive policy of utilizing the two priorities is proposed. GTSs are given to devices in a decreasing order of their priorities. Further, various scenarios of different data-based priorities and rate-based priorities are discussed. CAP and CFP traffic loads have also been taken into consideration. Our proposed scheme can satisfy the needs of time-critical and high-frequency devices. Details of the service differentiation mechanism and the GTS allocation mechanism are presented in the following sections.

\subsection{Service Differentiation Mechanism}
In the service differentiation mechanism, each device is adaptively assigned a data-based priority and a rate-based priority by the coordinator, according to the importance of data and past transmission feedback respectively. Assume there are $N$ devices in an 802.15.4-based PAN, and there are ~$N_d (0, 1, ..., N_{d-1})$~ data-based priority numbers and ~$N_r (0, 1, ..., N_{r-1})$~ rate-based priority numbers. Then the data-based priority number assigned to the device $n$ is defined as ~$P_{d_n}$~, and the rate-based priority number assigned to the device $n$ is defined as ~$P_{r_n}$~, where ~$0 \le P_{d_n} \le N_{d-1}$~ and ~$0 \le P_{r_n} \le N_{r-1}$~. In our ART-GAS, a larger priority number represents a higher priority for GTS allocation. Devices with higher data-based priorities are assumed to send time-critical data, such as alarms and emergency messages, and devices with higher rate-based priorities are considered to have more recent traffic and thus having higher probabilities to transmit data in the following superframe. The priority numbers of a device are internally maintained by the PAN coordinator.

\subsubsection{Data-based Priority}
In the MAC modifications, we set the total number of data-based priorities, ~$N_d$~ to 60 and classify all devices into three data-based priority levels according to whether there are exceptions or real-time requirements related to the data. The process is depicted by the state transition diagram in Fig. 2.

\begin{figure}[!t]
\setlength{\abovecaptionskip}{1pt}
\setlength{\belowcaptionskip}{5pt}
\renewcommand{\captionfont}{\bfseries}
\centering
\includegraphics[width=3.2in]{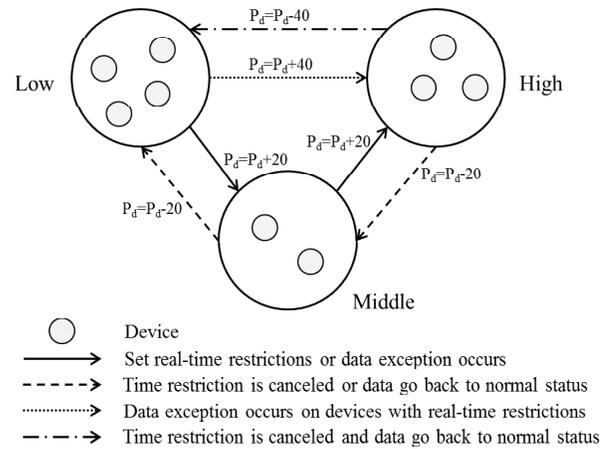}
\centering
\caption{State transition diagram}
\label{fig_sim}
\end{figure}

Each device has three states: Low, Middle and High, corresponding to its three levels respectively. State Low has the data-based priority numbers from 0 to 19, Middle has the numbers from 20 to 39, and High has the numbers from 40 to 59. They are defined as follows:

(1)	Low: All devices are placed in the Low state initially. There are neither transmission time restrictions nor exceptions in any of these data in this state. Their different data-based priority numbers are determined only by the importance of data without any other considerations.

(2)	Middle: Devices in the Middle state are sending data with real-time requirements or that certain exceptions related to the data occur during the current superframe. Whenever a device detects that an exception occurs or time restriction is set, it changes the state to Middle and modifies its data-based priority. When the real-time restrictions are canceled or the data value returns to the normal interval, the device will leave the Middle state for its original Low state.

(3)	High: High state means that there are certain exceptions in data with real-time limitations. This represents the highest level of data-based priority, which always indicates an emergency message or an alarm. In this case, we will set High state for the device and increase its data-based priority in order to privilege time-critical data transmissions. Similarly, whenever the conditions for High state cannot be satisfied, the state of the device will be changed and its data-based priority will be decreased.

\subsubsection{Rate-based Priority}
In addition to the data-based priorities, rate-based priorities are also dynamically assigned to each device by the PAN coordinator according to its recent transmission feedback. This method provides a good estimate of the future GTS usage behaviors of devices. Hence, GTS resources can be allocated to the needy devices with high frequencies of sending data, according to their different rate-based priorities. A waste of GTS resources is avoided in this way. Before presenting details of the rate-based priority assignment policy, we define CSMA/CA hit and CSMA/CA miss, GTS hit and GTS miss as follows:

(1)	CSMA/CA hit and CSMA/CA miss: If one device has attempted to access the channel in the CAP of the current superframe, the device is defined to have a CSMA/CA hit, no matter whether the attempt was successful or not. Otherwise, the device is considered to have a CSMA/CA miss.

(2)	GTS hit and GTS miss: If one device has issued a successful GTS request in the CAP or transmitted data within its allocated GTS to the PAN coordinator during the period of the current superframe, the device is defined to have a GTS hit. Otherwise, the device is considered to have a GTS miss [12].

It is easy to see that a CSMA/CA hit or GTS hit indicates more recent traffic for a certain device, whereas a CSMA/CA miss or GTS miss represents a comparatively light traffic. Hence, we can know about the recent data transmission behaviors of all devices through the occurrence of CSMA/CA hit/miss and GTS hit/miss. The rate-based priorities can be set dynamically according to the transmission feedback.

Specifically, whenever a CSMA/CA hit or GTS hit occurs on a device, the rate-based priority number will be increased by the PAN coordinator, and the priority of GTS allocation for the device upgrades. On the other hand, upon occurrence of a CSMA/CA miss or a GTS miss, the PAN coordinator decreases the rate-based priority of the device to reduce its opportunity for obtaining GTS resources. In this way, devices with more frequent transmissions will have larger probabilities to obtain GTS allocation in the subsequent superframe than devices with a light traffic.

Further, whenever a CSMA/CA hit/miss or a GTS hit/miss occurs, devices are not treated equally if they have different rate-based priorities. For example, devices with high rate-based priorities, which stay in a high traffic level, are more tolerated for temporarily unstable transmission behaviors. Such devices are slightly demoted to lower rate-based priorities upon occurrence of a CSMA/CA miss or GTS miss. Whereas, devices with comparatively low rate-based priorities are demoted more greatly for the same CSMA/CA miss or GTS miss. Additionally, when there is a CSMA/CA hit or a GTS hit, devices with lower rate-based priorities will be more greatly promoted to higher priorities to receive GTS service as soon as possible in order to avoid starvation of low-priority devices.

In the proposed scheme, devices with consecutive data transmissions are favored, and for a device that is idle for a period of time, its rate-based priority will be greatly degraded by the PAN coordinator and the unused GTSs will be reclaimed immediately for high traffic devices. Hence, a waste of GTS resources is avoided.

Assume device ~$i$~ maintains the following parameters at the beginning of the ~$t^{th}$~ superframe. ~$P_{{r_i}}^{t - 1}$~ and ~$P_{{r_i}}^t$~ are rate-based priorities for the previous superframe and current superframe respectively. ~$N_{CSMA/CA,hi{t_i}}^{t - 1}$~ , ~$N_{GTS,hi{t_i}}^{t - 1}$~ are the number of CSMA/CA hit and GTS hit that occurs on device ~$i$~ in the previous superframe. Assume that each device receives the beacon at the beginning of the current superframe, then the rate-based priority of device ~$i$~, ~$P_{{r_i}}^t$~will be updated as follows:
\begin{equation}\begin{array}{l}
P_{{r_i}}^t = P_{{r_i}}^{t - 1} - {M_{CSMA/CA}}(P_{{r_i}}^{t - 1}) - {M_{GTS}}(P_{{r_i}}^{t - 1})\\
            + {H_{CSMA/CA}}(P_{{r_i}}^{t - 1}) + {H_{GTS}}(P_{{r_i}}^{t - 1})
\end{array}\end{equation}

~${M_{CSMA/CA}}(P_{{r_i}}^{t - 1}), {M_{GTS}}(P_{{r_i}}^{t - 1}), {H_{CSMA/CA}}(P_{{r_i}}^{t - 1})$~, ~${H_{GTS}}(P_{{r_i}}^{t - 1})$~ can be determined from:

\begin{equation}{M_{CSMA/CA}}(P_{{r_i}}^{t - 1}) = \frac{{{\lambda _{CSMA/CA,miss}}}}{{P_{{r_i}}^{t - 1}}}\end{equation}
\begin{equation}{M_{GTS}}(P_{{r_i}}^{t - 1}) = \frac{{{\lambda _{GTS,miss}}}}{{P_{{r_i}}^{t - 1}}}\end{equation}
\begin{equation}{H_{CSMA/CA}}(P_{{r_i}}^{t - 1}) = \frac{{{\lambda _{CSMA/CA,hit}}}}{{P_{{r_i}}^{t - 1}}} \times {2^{N_{CSMA/CA,hi{t_i}}^{t - 1}}}\end{equation}
\begin{equation}{H_{GTS}}(P_{{r_i}}^{t - 1}) = \frac{{{\lambda _{GTS,hit}}}}{{P_{{r_i}}^{t - 1}}} \times {2^{N_{GTS,hi{t_i}}^{t - 1}}}\end{equation}
where ~${\lambda _{CSMA/CA,miss}}, {\lambda _{GTS,miss}}, {\lambda _{CSMA/CA,hit}}$~ and ~${\lambda _{GTS,hit}}$~ are constants, and ~${\lambda _{CSMA/CA,miss}}, {\lambda _{GTS,miss}}$~, $~{\lambda _{CSMA/CA,hit}}, {\lambda _{GTS,hit}}> $~ 0.

Fig. 3 presents the flowchart of the rate-based priority assignment process in a superframe.

\begin{figure}[!t]
\renewcommand{\captionfont}{\bfseries}
\centering
\includegraphics[width=3.2in]{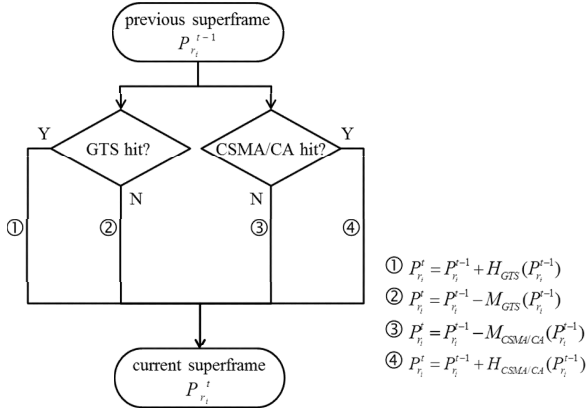}
\centering
\caption{Rate-based priority assignment process}
\label{fig_sim}
\end{figure}

\subsection{GTS Allocation Mechanism}
According to the IEEE 802.15.4 Specification [1], on receipt of a GTS request, the PAN coordinator shall first check if there is available capacity in the current superframe, based on the remaining length of the CAP and the desired length of the requested GTS. GTSs shall be allocated if the maximum number of GTSs (seven) has not been reached and allocating a GTS of the desired length would not reduce the length of the CAP to less than ~$aMinCAPLength$~.

Here, we describe the GTS allocation mechanism for our ART-GAS algorithm. The proposed mechanism will modify the FCFS GTS allocation policy in IEEE 802.15.4 standard, and optimize the passive deallocation scheme for GTS resources, which may result in starvation of light-traffic devices. In our GTS allocation mechanism, the GTS scheduling criteria are based on the priority numbers, the superframe length, and the GTS capacity of the superframe, compared with the original GTS mechanism. In this method, time-critical devices with higher frequencies of sending data are privileged over non time-critical and light-traffic devices. Details of the algorithm are presented as follows:

At first, we classify all devices into three states, Low, Middle and High, according to their data-based priority levels described in Section 3.1.1. For each state, we define a GTS scheduling criterion, which determines whether the GTS resources will be allocated to certain device. Criteria for different states are different, hence service for devices with Low, Middle and High data-based priority levels will be differentiated.

Let ~$P_i$~ denote priority of device ~$i$~ to obtain GTS allocation, and GTSs shall be given to devices in a decreasing order of their priorities. ~$P_L, P_M$~ and ~$P_H$~ denote the minimum priority required to be allocated GTS resources in state Low, Middle and High respectively. In other words, GTSs can be allocated to device ~$i$~ only when one of the following requirements is met:

(1)	Device ~$i$~ in Low state: ~${P_i} \ge {P_L}$~;

(2)	Device ~$i$~ in Middle state: ~${P_i} \ge {P_M}$~;

(3)	Device ~$i$~ in High state: ~${P_i} \ge {P_H}$~.

For devices in the High state, which is the highest level of data-based priorities, the real-time guarantee for emergency messages is the primary concern, whereas, frequencies of data transmissions are considered to be less important. In this scenario, we define ~$P_i$~ as ~${P_i} = {P_{{d_i}}}$~ and ~$P_H$~ as ~${P_H} = Min({P_i}) = 40$~. Hence, there will be no extra limitations for devices in the High state whenever they have data to transmit, provided there is available capacity in the current superfame.

For devices in the Middle state, both the data-based priority and rate-based priority are considered to determine a GTS allocation. Priority ~$P_i$~ is defined as~${P_i} = \sqrt {{P_{{d_i}}} \times {P_{{r_i}}}} $~, which is based on the device's data-based priority ~$P_d$~  and rate-based priority ~$P_r$~. Further, we define the threshold priority ~$P_M$~ as follows:

\begin{equation}{P_M} = \frac{{{\mu _M}\sum\limits_{i = 1}^N {\sqrt {P_i^2} } }}{{N{\Delta ^{BO}}}}\end{equation}
where ~${\mu _M}$~ and ~$\Delta $~ are both constants, and ~${\mu _M} > 0, 0 < \Delta  \le 1$~. ~$N$~ is the number of devices in the current IEEE 802.15.4-based PAN.

~$P_M$~ is presented here to filter unnecessary GTS allocations. It is dynamically adjusted and mainly depends on the ~$\frac{{\sum\limits_{i = 1}^N {\sqrt {P_i^2} } }}{N}$~ and ~${\Delta ^{BO}}$~ value. ~$BO$~ is an indication of CAP and CFP traffic load. As ~$BO$~ increases, there is a higher probability that a great number of devices have requested the GTS service in a superframe. Based on our service differentiation mechanism, the devices requesting GTS are assigned big priority numbers, even though they only have one request in the whole superframe. To prevent the scarce GTS resources from distributing to devices with extremely low frequency GTS requests in such a long superframe, a stricter threshold is needed. In this case, ~$P_M$~ will be set to a larger value to filter low priority devices. On the other hand, with a small ~$BO$~, the value of ~$P_M$~ can be decreased and the limitation for the device selection can be relaxed. ~$\frac{{\sum\limits_{i = 1}^N {\sqrt {P_i^2} } }}{N}$~ represents an average level of devices' priorities in a superframe. When most of the devices have low priorities, there is no need to allocate too many GTS resources for the devices. Too much dedicated bandwidth for GTS usage leads to resource wastage and to the degradation of the overall system performance. Instead, the GTS bandwidth should be transferred for contention-based accesses in CAP.

For devices in the Low state, rate-based priorities are considered to be more important since they all stay in low data-based priority levels. In this case, ~$P_i$~ is defined as ~${P_i} = {P_{{r_i}}}$~. Similar to the Middle state, the threshold value ~$P_L$~ is defined as follows:

\begin{equation}{P_L} = \frac{{{\mu _L}\sum\limits_{i = 1}^N {\sqrt {{P_i}^2} } }}{{N{\Delta ^{BO}}}}\end{equation}

However, this is a more restricted limitation for GTS allocation compared with the Middle state since we have different ~${\mu _M}$~ and ~${\mu _L}$~ values.

\section{Performance Evaluation}
\subsection{Simulation Setup}
In this section, we present a simulation study based on an accurate model of IEEE 802.15.4 using OMNET++ simulator, and assess the performance of the proposed ART-GAS algorithm.

In the simulation, we use a star topology with single PAN coordinator and 20 devices deployed in the area of 1000mm*1000mm. All GTS transmissions are assumed to be successful. Each device is allocated at most one GTS slot, and according to the IEEE 802.15.4 Specifications, the maximum GTS number in a superframe is seven. If there are no sufficient GTS resources for the request, the device will reissue the request for the packet in the subsequent superframe.

Additionally, two traffic types generated by devices are considered: heavy traffic and light traffic. ~${\chi _h}$~ and ~${\chi _i}$~ represent respectively the interarrival rates for the heavy-traffic and light-traffic devices. In the simulations, we have ~${\chi _h} = 0.35/s$~ and ~${\chi _i} = 0.15/s$~. Such rate settings are reasonable in 802.15.4-based PANs, since IEEE 802.15.4 targets low-rate wireless communications. Further, let ~$N_h$~ denote the number of heavy-traffic devices. Thus, the GTS traffic load will be ~$\Gamma  = {N_h}{\chi _h} + (N - {N_h}){\chi _l}$~. Table 1 lists the input parameters for our simulation model.

\begin{table}[!t]
\setlength{\abovecaptionskip}{0pt}
\setlength{\belowcaptionskip}{5pt}
\renewcommand{\captionfont}{\bfseries}
\centering
\caption{Parameters for simulation model}
\begin{tabular}{|c|c|}
\hline 
    Parameters    &   Value    \\
\hline
Network topology&Star topology\\
\cline{1-2}
Number of devices&20\\
\cline{1-2}
Frame size&127B\\
\cline{1-2}
Transmission rate&200kps\\
\cline{1-2}
BO(SO)&3\\
\cline{1-2}
Buffer size&150\\
\cline{1-2}
~${\chi _i}$~&0.15/s\\
\cline{1-2}
~${\chi _h}$~&0.35/s\\
\hline
\end{tabular}
\end{table}

We develop a simulation model-Path Loss Model to investigate the performance of our ART-GAS algorithm. Based on [15], the path loss model is suitable for both narrow band and Ultra-wide Bandwidth band. We assume that the human body is 171cm high and his weight is 63kg on average. The path loss with 400MHz can be calculated as the following formula:

\begin{equation}PL(d)\left[ {dB} \right] = a\lg d + b + c + N\end{equation}

With the assumption ~$PL(d,f)$~ [16] for body surface to body surface propagation at distance d (150mm $<$ d $<$ 1000mm) and frequency f = 2.4GHz (400MHz $<$ f $<$ 2500GHz), the following formula can be obtained from Eqs. (8):

\begin{equation}PL(d,f)\left[ {dB} \right] =  - 27.6\lg d - 46.5\lg f + 157 + Q\end{equation}

where $Q$ is the shadowing component which follows log-normal distribution with standard deviation 4.12dB.

The performance of devices with different priorities is analyzed in terms of average delay and probability of success, which reflects the adaptive and real-time guarantee for high priority devices achieved by the network. The traffic load represents all command and data frames generated by the MAC layers of 20 devices.

We consider five different scenarios, presented in Table 2. Each scenario is simulated with our ART-GAS algorithm.

\begin{table}[!t]
\renewcommand{\captionfont}{\bfseries}
\setlength{\abovecaptionskip}{0pt}
\setlength{\belowcaptionskip}{5pt}
\centering
\caption{Simulation scenarios} 
\begin{tabular}{|c|c|c|}
\hline 
Scenario&Data-based priority&Rate-based priority\\
\hline
Scenario 1&5&10\\
\cline{1-3}
Scenario 2&10&10\\
\cline{1-3}
Scenario 3&25&15\\
\cline{1-3}
Scenario 4&25&20\\
\cline{1-3}
Scenario 5&50&10\\
\hline
\end{tabular}
\end{table}

In addition, we will compare the proposed algorithm with the GTS allocation mechanism specified in IEEE 802.15.4, in terms of bandwidth utilization and average waiting time. Details and discussions of the simulation results are presented in the following section.

\subsection{Results and Analysis}
The following performance measures are considered: success probability (S), average delay (D), average waiting time (W) and bandwidth utilization (B). Success probability is computed as network throughput ~$Th$~ divided by the traffic load ~$\Gamma, S = Th/\Gamma$~. It reflects the degree of reliability achieved by the network for successful transmissions. We denote by ~$S(\Gamma )$~ the success probability as a function of the traffic load ~$\Gamma$~. D is the average delay experienced by a data frame from the start of its generation by the application layer to the end of its reception by the analyzer. We denote by ~$D(\Gamma)$~ the average delay as a function of the offered load ~$\Gamma$~. The average packet waiting time and bandwidth utilization are also important metrics for our proposed ART-GAS algorithm, which are presented to assess performance of our ART-GAS and the GTS mechanism specified in IEEE 802.15.4 standard.

Fig. 4 clearly shows the impact of the proposed service differentiation mechanism of ART-GAS, related to different priorities on the success probability. As it was intuitively expected, setting different data-based priorities and rate-based priorities results in higher throughputs for high priority devices due to the privileges given to them. On the other hand, devices with low priorities have relatively lower throughputs and smaller probabilities of success. This provides insurance of reliable transmission for the time-critical and high-traffic devices.

\begin{figure}[!t]
\renewcommand{\captionfont}{\bfseries}
\centering
\includegraphics[width=3.2in]{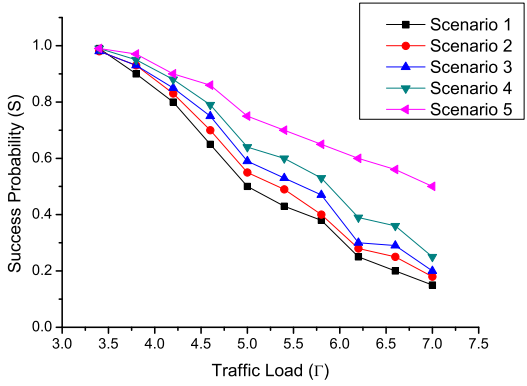}
\centering
\caption{Success probability of devices with different priorities}
\label{fig_sim}
\end{figure}

In Fig. 5, it is obvious that devices with lower priorities have greater average delays. This is because low priority devices have a small probability of obtaining GTS resources than high priority devices.

\begin{figure}[!t]
\renewcommand{\captionfont}{\bfseries}
\centering
\includegraphics[width=3.2in]{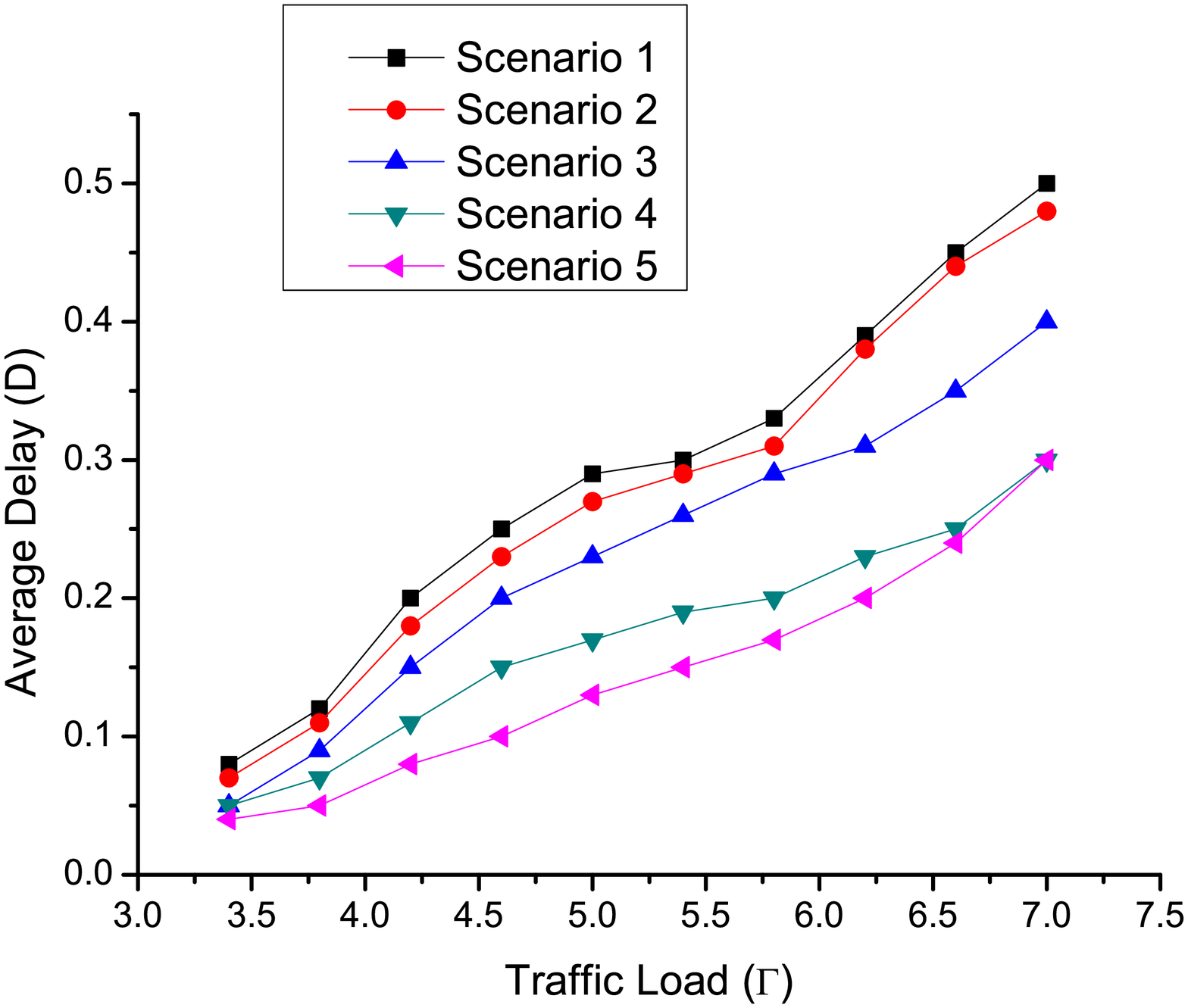}
\centering
\caption{Average delay of devices with different priorities}
\label{fig_sim}
\end{figure}

Further, we provide a performance comparison of our ART-GAS algorithm and the original GTS mechanism in terms of average waiting time and bandwidth utilization. In Fig. 6, we can observe the increasing rate of the latency for our ART-GAS algorithm is much smaller under all traffic distributions as traffic load increases. The proposed scheme provides more resistance to the increase in the traffic load, even if the network size is large. On the other hand, for the original IEEE 802.15.4 GTS allocation mechanism, traffic load has a greater impact on the latency. As traffic load increases, the average packet waiting time will significantly increase. The increase in latency results from the inflexibility of GTS allocation in the IEEE 802.15.4 Specifications. In this case, most of the GTS resources are occupied by heavy-traffic devices for a long time, which may lead to the starvation of light traffic devices.

\begin{figure}[!t]
\renewcommand{\captionfont}{\bfseries}
\centering
\includegraphics[width=3.2in]{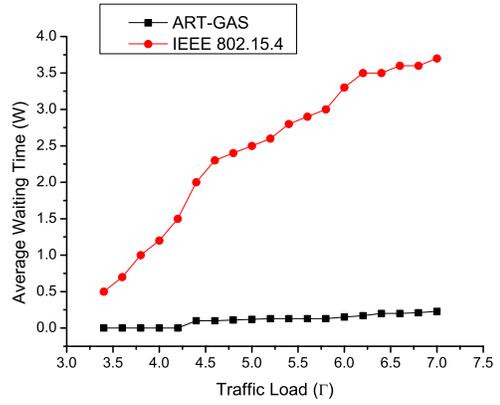}
\centering
\caption{Effect of traffic load on average waiting time}
\label{fig_sim}
\end{figure}

\begin{figure}[!t]
\renewcommand{\captionfont}{\bfseries}
\centering
\includegraphics[width=3.2in]{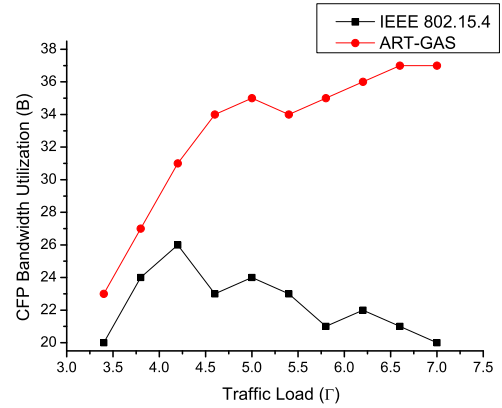}
\centering
\caption{Effect of traffic load on CFP bandwidth utilization}
\label{fig_sim}
\end{figure}

Fig. 7 shows that the proposed scheme has better bandwidth utilization than IEEE 802.15.4. Compared with the original GTS allocation mechanism, our ART-GAS algorithm keeps a stable even increasing value of bandwidth utilization as traffic load increases. This is because the proposed scheme provides a service differentiation mechanism for heavy traffic and light traffic devices, and hence GTS resources can be fully used by data transmissions. Furthermore, the passive GTS deallocation mechanism is optimized in the ART-GAS. The allocated but unused GTS can be declaimed as soon as possible for new data transmissions.

\section{Conclusions}
In this paper, we have proposed an adaptive and real-time GTS allocation scheme (ART-GAS) based on two mechanisms. In the first mechanism, we present a service differentiation algorithm that dynamically assigns data-based priorities and rate-based priorities to all devices. Higher priorities are distributed to time-critical and heavy-traffic devices. The second mechanism of ART-GAS allocates GTS resources to devices according to their priorities assigned previously, threshold priorities are defined to filter unnecessary allocations, thus bandwidth wastage can be avoided. Our proposed scheme can be implemented in the IEEE 802.15.4 MAC standard without adding any new message type. An analytic model was developed to evaluate the performance of IEEE 802.15.4 and ART-GAS, which has been validated against simulation experiments. Simulation results demonstrate that the proposed scheme significantly improves network performance in terms of average delay, success probability, average waiting time and bandwidth utilization.


\section{Acknowledgments}
This work was partially supported by the Natural Science Foundation of China under Grant No. 60903153, the Fundamental Research Funds for Central Universities (DUT10ZD110), the SRF for ROCS, SEM, and DUT Graduate School (JP201006).


\end{document}